\documentclass[conference]{IEEEtran}
\IEEEoverridecommandlockouts
\usepackage{color}
\usepackage{amsmath}
\usepackage{amssymb}
\usepackage{graphicx}
\usepackage{subfigure}
\usepackage{psfrag}
\usepackage{cite}
\usepackage{amsfonts}
\usepackage{epsfig}
\usepackage{enumerate}

\usepackage[
top    = 0.753in,
bottom = 0.98in,
left   = 0.673in,
right  = 0.673in]{geometry}

\def\b0{{\boldsymbol 0}}

\IEEEoverridecommandlockouts 

\usepackage[
top    = 0.753in,
bottom = 0.98in,
left   = 0.673in,
right  = 0.673in]{geometry}

\def\BibTeX{{\rm B\kern-.08em{\sc i\kern-.025em b}\kern-.08em
    T\kern-.1667em\lower.7ex\hbox{E}\kern-.125emX}}

\begin{document}
\title{Capacity-Net-Based RIS Precoding Design without Channel Estimation for mmWave MIMO System
\thanks{This work was supported in part by the Academia Sinica (AS) under Grant 235g Postdoctoral Scholar Program, and in part by the National Science and Technology Council (NSTC) of Taiwan under Grant 112-2218-E-110-004, 112-2218-E-110-003.}}

\author{\IEEEauthorblockN{Chun-Yuan Huang$^{1}$, Po-Heng Chou$^{2}$, Wan-Jen Huang$^{1}$, Ying-Ren Chien$^{3}$, and Yu Tsao$^{2}$\\}
\IEEEauthorblockA{$^{1}$Institute of Communication Engineering (ICE), National Sun Yat-sen University (NSYSU), Kaohsiung, 80424, Taiwan\\
$^{2}$Research Center for Information Technology Innovation (CITI), Academia Sinica (AS), Taipei, 11529, Taiwan\\
$^{3}$Department of Electrical Engineering, National Ilan University (NIU), I-Lan, 26041, Taiwan\\
E-mails: m113070012@nsysu.edu.tw, d00942015@ntu.edu.tw, wjhuang@faculty.nsysu.edu.tw,\\ yrchien@niu.edu.tw, yu.tsao@citi.sinica.edu.tw}
}

\maketitle

\begin{abstract}
In this paper, we propose Capacity-Net, a novel unsupervised learning approach aimed at maximizing the achievable rate in reflecting intelligent surface (RIS)-aided millimeter-wave (mmWave) multiple input multiple output (MIMO) systems. To combat severe channel fading of the mmWave spectrum, we optimize the phase-shifting factors of the reflective elements in the RIS to enhance the achievable rate. However, most optimization algorithms rely heavily on complete and accurate channel state information (CSI), which is often challenging to acquire since the RIS is mostly composed of passive components. To circumvent this challenge, we leverage unsupervised learning techniques with implicit CSI provided by the received pilot signals. Specifically, it usually requires perfect CSI to evaluate the achievable rate as a performance metric of the current optimization result of the unsupervised learning method. Instead of channel estimation, the Capacity-Net is proposed to establish a mapping among the received pilot signals, optimized RIS phase shifts, and the resultant achievable rates.
Simulation results demonstrate the superiority of the proposed Capacity-Net-based unsupervised learning approach over learning methods based on traditional channel estimation. 
\end{abstract}

\begin{IEEEkeywords}
Millimeter wave (mmWave), reflecting intelligent surface (RIS), deep learning (DL), multiple-input multiple-output (MIMO), implicit channel estimation, Capacity-Net.
\end{IEEEkeywords}

\section{Introduction}

Millimeter-wave (mmWave) bands, spanning 24–52 GHz, are extensively utilized to expand the spectrum for offloading vast amounts of data in next-generation mobile networks~\cite{Haider2022}. The transmission across mmWave bands offers high-speed service owing to the ultra-wide vacant spectrum. However, the shorter wavelength of mmWave makes its signal energy prone to attenuation by environmental obstacles. To address this challenge, multi-input-multi-output (MIMO) assisted by \emph{Reflecting Intelligent Surfaces} (RIS) has been proposed to mitigate energy attenuation in mmWave transmission and improve the coverage especially when the direct path between the transmitter and receiver is blocked \cite{Wu2020, Wu2021, RIS-implicitCSI, RIS-implicitCSI2}. RIS comprises multiple passive elements and an FPGA controller capable of adjusting the phase shifting of the incident signals. By employing an appropriately designed phase shift of the RIS, the reflected signal can significantly improve the spectral and energy efficiency.

Various methods were proposed to maximize the achievable rate of the RIS-assisted MIMO system, including the deep learning (DL) based methodse~\cite{Song, Zhang, DRL}, and most of them were developed based on perfect channel state information (CSI). Some methods even demand the estimation of individual links from the transmitter to the RIS and from the RIS to the receiver. Unfortunately, it is intractable to estimate the individual channel matrices of the links connected to the RIS, because most reflective elements are passive. Moreover, RIS typically comprises a large number of reflective elements, resulting in a substantial quantity of channel parameters, which is costly to estimate all channel parameters. Therefore, we propose a novel unsupervised learning approach to optimize the system based on implicit channel estimation~\cite{RIS-implicitCSI, RIS-implicitCSI2, Spatial-DL}, by skipping traditional pilot-based channel estimation. During the training phase, the proposed unsupervised learning method leverages the indirect mapping between the phase shifts of RIS and the received pilot signal to maximize the achievable rate, instead of relying on explicit labels.

We adopt the negative value of the achievable rate as a loss function to evaluate the performance of unsupervised learning.
However, it usually requires perfect CSI to assess the achievable rate in terms of the RIS parameters.
To tackle this challenge, we introduce an additional neural network dubbed Capacity-Net, drawing inspiration from Quality-Net~\cite{Quality-Net, Quality-Net2} and STOI-Net~\cite{STOI-Net, STOI-Net2}. Quality-Net and STOI-Net, both incorporating an extra neural network, approximates performance measurements of quality and intelligibility scores in speech enhancement applications. 
At first, DL-based speech enhancement models rely on supervised learning, necessitating pairs of clean and noisy speech during training. Subsequently, various studies have explored unsupervised learning methods for speech enhancement to alleviate the need for paired clean-noisy speech~\cite{Bie2021, Xiang2020, CycleGAN, Fujimura2021}. Nonetheless, these approaches still necessitate either clean or noisy speech for training. Hence, non-intrusive quality assessment mechanisms have been proposed in~\cite{Quality-Net, Quality-Net2, STOI-Net, STOI-Net2}, where no clean reference speech is required. Traditionally, mean squared error (MSE) serves as the standard objective function for optimizing DL model parameters. However, MSE might not accurately capture human auditory perception, leading to suboptimal performance. Two human-perception-related metrics, perceptual evaluation of speech quality (PESQ) \cite{PESQ}, and short-time objective intelligibility (STOI)~\cite{STOI}, have demonstrated high correlation with quality and intelligibility scores rated by human listeners, respectively. Due to its highly complex and non-differentiable nature, PESQ is not directly suitable for optimizing DL models. In our previous studies~\cite{Quality-Net, Quality-Net2}, Quality-Net was introduced to optimize enhancement models using an approximated PESQ function, which is differentiable, and to extract features from the training data. Quality-Net has exhibited remarkable performance in evaluating noisy and processed speech without the need for clean speech as a reference.



In this paper, we introduce Capacity-Net as an alternative to traditional metric functions like Quality-Net or STOI-Net in unsupervised learning for computing achievable rates. To eliminate the need for CSI during training, we employ supervised learning to craft the Capacity-Net neural network model. This model learns to map received pilot signals, optimized Reconfigurable Intelligent Surface (RIS) phase shifts, and resulting achievable rates without relying on CSI. Our key contributions are outlined as follows:
\begin{itemize}
\item During the online inference phase, the proposed Capacity-Net-based unsupervised learning directly approximates the achievable rate function. It establishes the mapping between the optimized RIS precoder and pilot signal without requiring explicit labels.
\item During the offline training phase, only the received pilot signal is necessary for training the proposed Capacity-Net-based unsupervised learning. There's no reliance on perfect CSI from prior estimation.
\item Simulation results demonstrate the superiority of the proposed Capacity-Net-based unsupervised learning over other frameworks. It showcases robustness against channel variation.
\item Despite still requiring perfect CSI during training, Capacity-Net exhibits inherent generalizability. It eliminates the need for re-training when encountering slight channel variations.
\end{itemize}

\section{System and Channel Models}

\begin{figure}
\begin{center}
\includegraphics[width = 80mm]{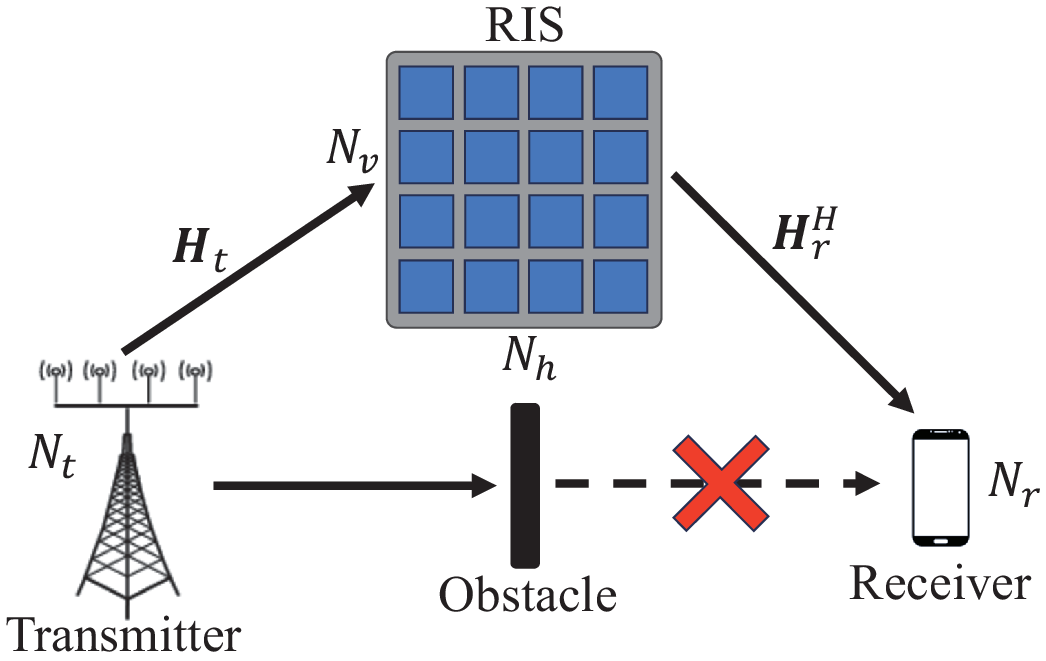}
\end{center}
\vspace{-0.1in}
\caption[c]{The RIS-aided mmWave MIMO system.}
\vspace{-0.2in}
\label{fig_system_model}
\end{figure}

Consider a RIS-aided MIMO system, as shown in Fig.~\ref{fig_system_model}, where the transmitter and receiver are equipped with 
uniform linear arrays (ULAs) with 
$N_t$ and $N_r$ elements, respectively. The direct path between the transmitter and receiver is obstructed. The reflective elements of RIS are configured as a uniform planar array (UPA) with $N = N_{h} \times N_{v}$ passive components, where $N_h$ and $N_v$ are the numbers of reflective elements in horizontal and vertical directions, respectively. Let $\boldsymbol{s}(\ell)\in\mathbb{C}^{N_t \times M}$ be the space-time block codeword (STBC)~\cite{STBC} transmitted during the $\ell$-th time-slot, where $M$ is the block length. The STBC codeword shall satisfy the total power constraint of transmitted power $P_T$.
After reflected by the RIS, the received pilot signal in the $\ell$-th time slot is written as
\begin{equation}
\boldsymbol{r}(\ell) = \boldsymbol{H}_r^H \mathrm{diag}(\boldsymbol{v}(\ell)) \boldsymbol{H}_t \boldsymbol{s}(\ell) + \boldsymbol{w}(\ell), 
\end{equation}
where $\boldsymbol{H}_t \in \mathbb{C}^{N \times N_t}$ and $\boldsymbol{H}_r^H \in \mathbb{C}^{N_r \times N}$ are the channel matrices of the transmitter to the RIS and the RIS to the receiver, respectively. $\textrm{diag} (\cdot)$ is vector-to-diagonal matrix operator, $\boldsymbol{w}\in \mathbb{C}^{N_r \times M}$ is the additive white Gaussian noise (AWGN) matrix with i.i.d entries being $CN(0, \sigma^2)$. The RIS re-shape the MIMO channel by adjusting the phase-shifting of the incident signals, denoted by the response vector of the RIS  $\boldsymbol{v}(\ell) = [e^{j\theta_{1, n}}, e^{j\theta_{2, n}}, \ldots, e^{j\theta_{N, n}}]$, where $\theta_i \in [0, 2\pi)$ denotes the phase shift of the $i$-th reflective element.


The Rician fading channel is assumed for the mmWave transmission of both the links from the transmitter to the RIS and from the RIS to the receiver~\cite{mmWave3}.
The channel of the transmitter-RIS and the RIS-receiver links are composed of Line-of-sight (LOS) and non-line-of-sight (NLOS) paths, which can be respectively written by
\begin{equation}
\boldsymbol{H}_t = \sqrt{\frac{L_{t} K_t}{K_t + 1}} \hat{\boldsymbol{H}}_t + \sqrt{\frac{L_{t}}{K_t + 1}} \tilde{\boldsymbol{H}}_t,
\label{channel_S_RIS}
\end{equation}
\begin{equation}
\boldsymbol{H}_r^H = \sqrt{\frac{L_{r} K_r}{K_r + 1}} \hat{\boldsymbol{H}}_r^H + \sqrt{\frac{L_{r}}{K_r + 1}} \tilde{\boldsymbol{H}}_r^H,
\label{channel_RIS_D}
\end{equation}
where $\hat{\boldsymbol{H}}_t$ and $\tilde{\boldsymbol{H}}_t$ denote the LOS and NLOS components of the transmitter-RIS link,  $\hat{\boldsymbol{H}}_r^H$ and $\tilde{\boldsymbol{H}}_r^H$ are the LOS and NLOS components of the RIS-receiver link, $L_{t}$ and $L_{r}$ are the path losses of the transmitter-RIS link and the RIS-receiver link, respectively. In addition, $K_t$ and $K_r$ are the Rician factors for the transmitter-RIS and RIS-receiver links, respectively.
Without loss of generality, we assume that the spacings of transmit antennas, receive antenna, and the RIS reflective components are all equal to half of the wavelength. Denote an $N$-dimensional steering  vector as $\boldsymbol{a}_N(\psi)=[1, e^{j\pi\psi},e^{j2\pi\psi},\cdots,e^{j(N-1)\pi\psi} ]$.
The LOS components of two channel matrices are given by
\begin{align*}
\hat{\boldsymbol{H}}_{t}&=\Big[{\boldsymbol{a}}_{N_h}^H(\cos\phi_{A,0}\sin\theta_{A,0})\otimes{\boldsymbol{a}}_{N_v}^H(\sin\phi_{A,0})\Big]{\boldsymbol{a}}_{N_t}(\sin\theta_{D,0}^{(t)}),\\
\hat{\boldsymbol{H}}_{r}^H&={\boldsymbol{a}}_{N_r}^H(\sin\theta_{A,0}^{(r)})\Big[{\boldsymbol{a}}_{N_v}(\sin\phi_{D,0})\otimes{\boldsymbol{a}}_{N_h}(\cos\phi_{D,0}\sin\theta_{D,0})\Big],
\end{align*}
where $\otimes$ is the Kronecker product of two vectors, $\theta_{D,0}^{(t)}$ is the angle of departure (AoD) of the LOS path at the transmitter,
$\theta_{A,0}$ and $\phi_{A,0}$ are respectively the azimuth and elevation
angles of arrival (AoA) of the LOS path at the RIS, $\theta_{D,0}$ and $\phi_{D,0}$ are respectively the azimuth and elevation AoD of the LOS at the RIS, and $\theta_{A,0}^{(r)}$ is the AoA of the LOS at the receiver. On the other hand, the non-LOS components of two channel matrices
are given by
\begin{align*}
\tilde{\boldsymbol{H}}_{t}\!=\!\!\sum_{p=1}^{\Lambda_t}& z_{t, p}\Big[{\boldsymbol{a}}_{N_h}^H(\cos\phi_{A,p}\sin\theta_{A,p}) \!\otimes\!{\boldsymbol{a}}_{N_v}^H(\sin\phi_{A,p})\Big]\times\\
&{\boldsymbol{a}}_{N_t}(\sin\theta_{D,p}^{(t)}),\\
\tilde{\boldsymbol{H}}_{r}^H\!=\!\!\sum_{p=1}^{\Lambda_r}& z_{r, p}{\boldsymbol{a}}_{N_r}^H(\sin\theta_{A,p}^{(r)})\times
\Big[{\boldsymbol{a}}_{N_v}(\sin\phi_{D,p})\!\\
&\otimes\!{\boldsymbol{a}}_{N_h}(\cos\phi_{D,p}\sin\theta_{D,p})\Big],
\end{align*}
where $\Lambda_t$ and $\Lambda_r$ are the numbers of non-LOS paths of two links, and $z_{t, p}\sim{\cal CN}(0,\frac{1}{\Lambda_t})$ and $z_{r, p}\sim{\cal CN}(0,\frac{1}{\Lambda_r})$ are the complex gains of the $p$-th non-LOS paths. The definitions of the AoAs and AoDs of both links are identical for the LOS and non-LOS components.

Define $\boldsymbol{H}_{\rm eff} = \boldsymbol{H}_r^H \mathrm{diag}(\boldsymbol{v}) \boldsymbol{H}_t$ as the equivalent channel between transmitter and receiver.
At the receiver, the achievable rate is expressed as
\begin{equation}
\label{rate}
R(\boldsymbol{v}) = \log_2 \det \left( \boldsymbol{I} + \frac{P_T}{\sigma^2 N_t} \boldsymbol{H}_{\text{eff}}^H \boldsymbol{H}_{\text{eff}} \right).
\end{equation}

\section{Problem Formulation}

The purpose is to design the vector of the RIS phase shifts $\boldsymbol{v}$ without perfect channel information to maximize the achievable rate $R(\boldsymbol{v})$. We adopt an unsupervised neural network model $g(\cdot)$ to generate the optimal RIS phase shift vector $\boldsymbol{v}_{\rm opt}$ that can maximize the achievable rate based on the matrix of pilot signal $\boldsymbol{Y}\triangleq [{\boldsymbol{r}}(1), {\boldsymbol{r}}(2), \cdots, {\boldsymbol{r}}(L)]$, where $L$ is the number of pilot codewords.

Thus, each set of pilot signals corresponds to a specific RIS phase shifting and is used to provide implicit information on the current channel status. 
To maximize the achievable rate, we formulate an optimization problem in the following:
\begin{equation}
\label{problem_formulation}
\begin{aligned}
& \max_{\boldsymbol{v}=g(\boldsymbol{Y})}   R(\boldsymbol{v}),\\
\text{s.t} \quad  
& \quad \lvert v_i \rvert = 1, \quad i = 1, 2, \ldots, N.\\
\end{aligned}
\end{equation}
where the constraint limits the maximum transmission power and normalizes the amplitude of each reflective element $ \left| v_i \right| = 1, \forall i$.
The optimization problem is non-convex because of the constraint on the RIS reflective matrix. 
To solve this intractable problem, we may optimize the RIS precoder by exhaustive search. However, exhaustive search introduces significant computational complexity and lengthy processing times because of its iterative process, and it may yield sub-optimal solutions. In our previous work, the dimension-wise sinusoidal maximization (DSM) algorithm~\cite{DSM} needs to use CSI data for iterative computation until convergence.
Hence, we further propose a DL approach that treats the pilot signals as states, and leverages neural networks to find the sub-optimal RIS precoder $\boldsymbol{v}_{\rm opt}$.

To facilitate the precoding design, the RIS phase-shifting factor shall be properly designed to fully explore the channel status. Hence, the phase-shifting parameters in the $\ell$-th time slot $\mathrm{diag}(\boldsymbol{v}(\ell))$ are chosen from a predefined codebook. Since the channel matrices are composed of beam-steering vectors in our study, the beam-steering vectors are used to construct the codebook with quantized angles. Specifically, the codebook $\mathcal{C}_{\rm b}$ is constructed as follows,
\begin{align}
   \mathcal{C}_{\rm b}= \Big\{&\boldsymbol{a}_{N_{h}}(\theta_{h,i}) \otimes \boldsymbol{a}_{N_{v}}(\theta_{v,j}),\notag\\ 
   &i=1,2,\ldots, N_h, j=1,2,\ldots, N_v\Big\},
\end{align}
where $\theta_{h,i}=\frac{2(i-1)}{N_{h}}$ and $\theta_{v,j}=\frac{2(j-1)}{N_{v}}$.  

\section{Capacity-Net-Based Unsupervised Learning}

\begin{figure}[t]
\begin{center}
\includegraphics[width = 90mm]{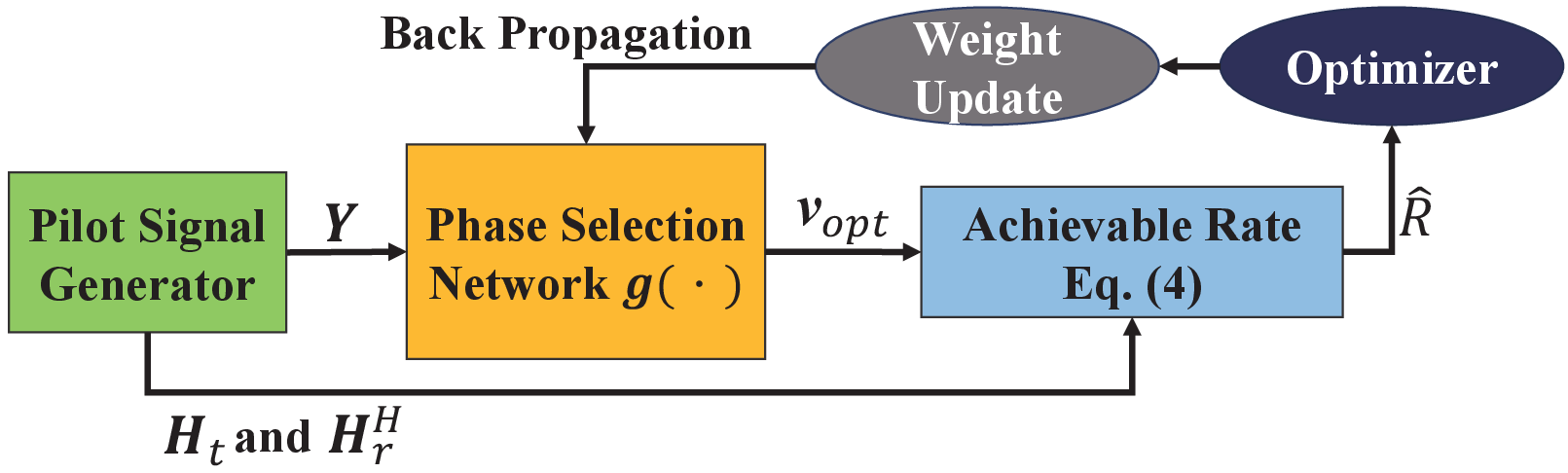}
\end{center}
\vspace{-0.1in}
\caption[c]{Unsupervised learning model training flow chart.}
\vspace{-0.1in}
\label{fig.Unsupervised}
\end{figure}
\begin{figure}[t]
\begin{center}
\includegraphics[width = 90mm]{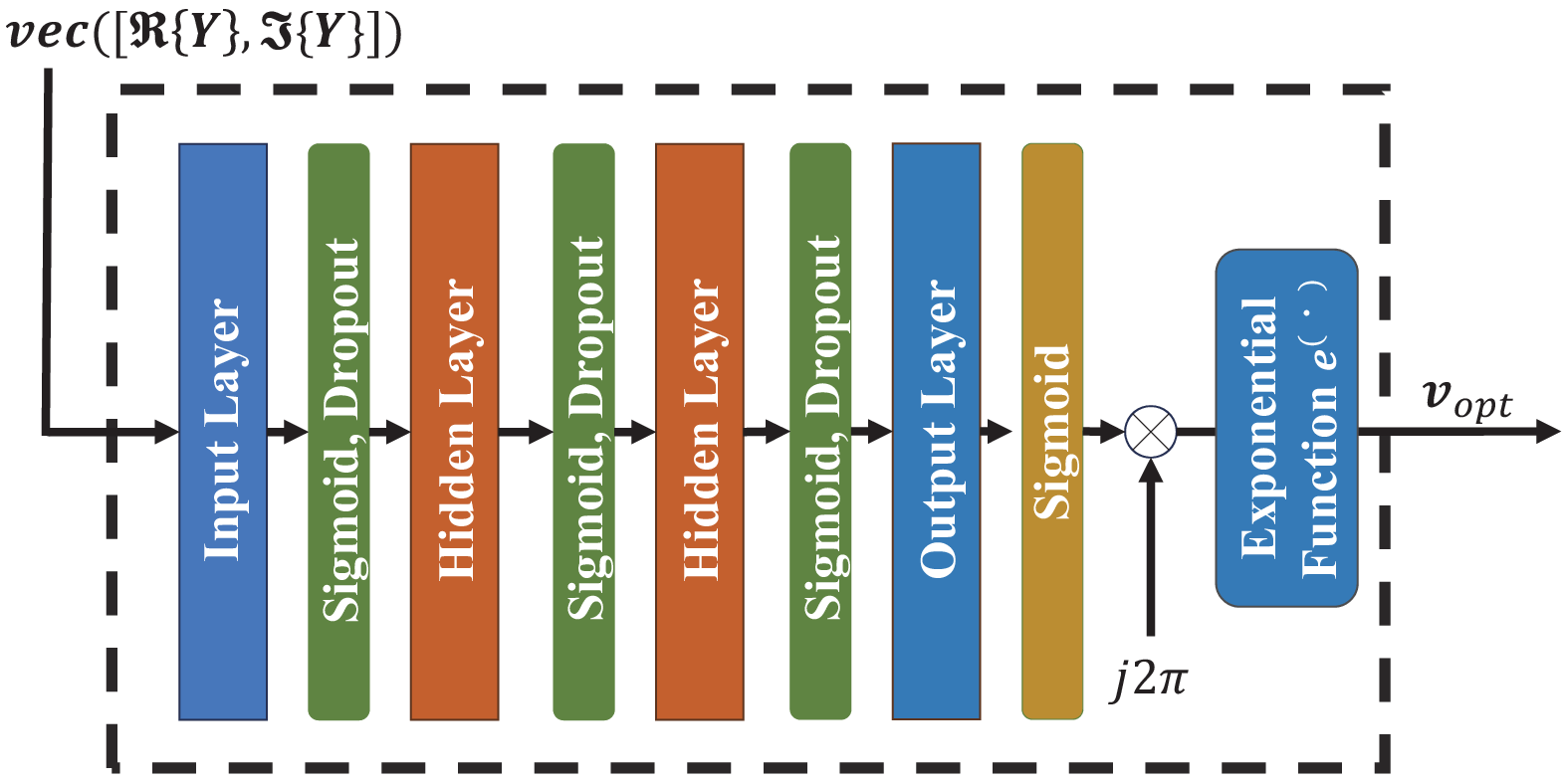}
\end{center}
\vspace{-0.1in}
\caption[c]{Phase selection neural network architecture.}
\vspace{-0.2in}
\label{fig.neural_network}
\end{figure}

To design the generative function of RIS precoder $g(\cdot)$ in Eq.~(\ref{problem_formulation}), we adopted unsupervised learning for training a neural network. During the training phase of unsupervised learning, the neural network does not depend on explicit labels but instead relies on the implicit information of CSI data to guide learning.

\subsection{Unsupervised Learning}

 As shown in Fig.~\ref{fig.Unsupervised}, we use the known generative pilot signal $\boldsymbol{Y}$, the phase selection neural network $g(\cdot)$, and the perfect CSI data $\boldsymbol{H}_t$, $\boldsymbol{H}_r^H$ for designing the RIS precoder $\boldsymbol{v}_{\rm opt}$ to achieve the maximize achievable rate $R(\boldsymbol{v}_{\rm opt})$ in Eq.~(\ref{rate}).
The phase shift selection neural network $g(\cdot)$ is illustrated in Fig.~\ref{fig.neural_network}. It consists of a fully connected neural network architecture. The input data of this neural network comprises the received pilot signals $\boldsymbol{Y}$ which are segregated into the real and imaginary parts and flattened into a one-dimensional vector. The goal of the neural network is to learn a mapping that transforms this one-dimensional vector into a phase vector $\boldsymbol{v}$ to maximize the achievable rate.

\subsection{Capacity-Net}

\begin{figure}[t]
\begin{center}
\includegraphics[width = 80mm]{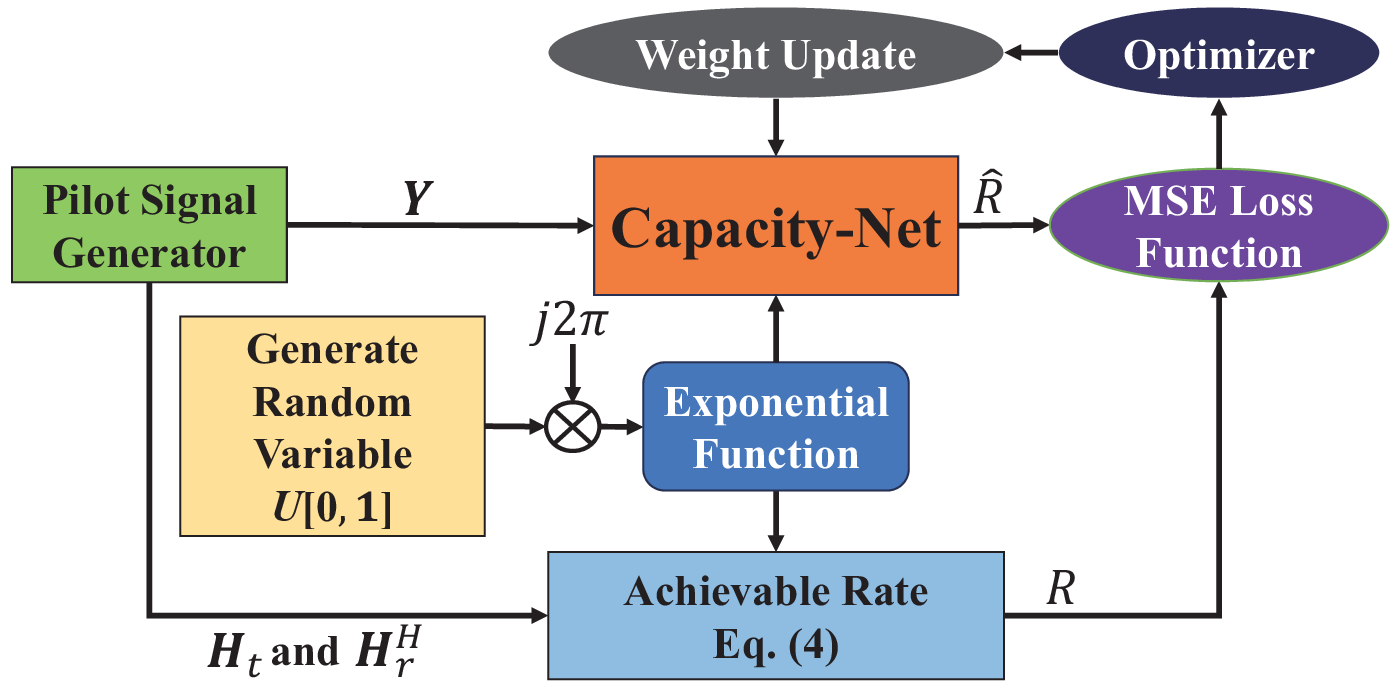}
\end{center}
\caption[c]{Capacity-Net model training flow chart.}
\label{fig.Capacity-Net}
\end{figure}
\begin{figure}[t]
\begin{center}
\includegraphics[width = 70mm]{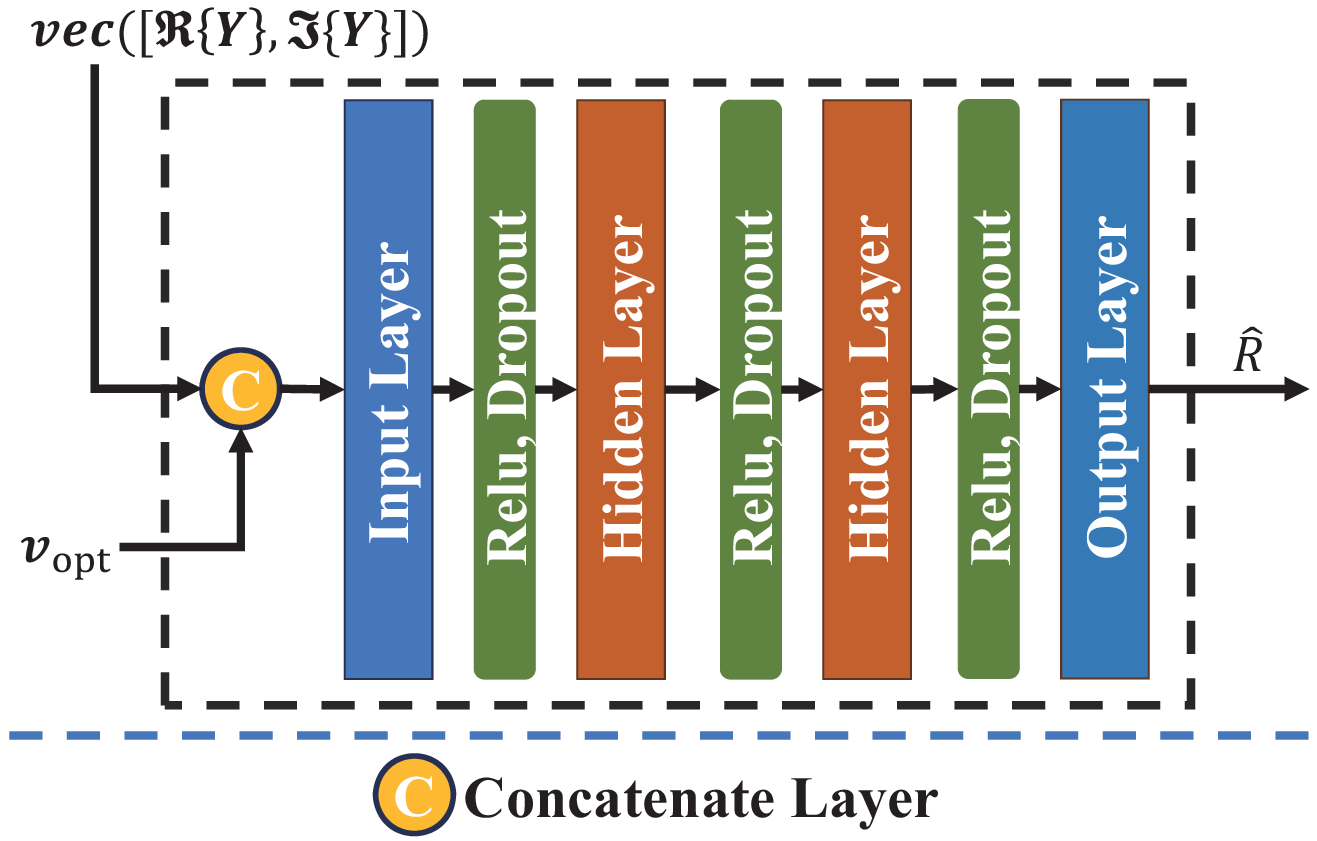}
\end{center}
\caption[c]{Capacity-Net neural network architecture.}
\label{fig.Capacity-Net_neural}
\end{figure}

Since it is impractical to obtain perfect CSI, the potential channel variations lead to a severe challenge to the channel estimation between transmitter-RIS and RIS-receiver. Our purpose is to train the neural network to design the RIS phase shift vector through the reception of pilot signals without channel estimation. To address this issue, we propose a novel concept of Capacity-Net. The proposed Capacity-Net is a pre-trained neural network to find the RIS precoder to maximize the achievable rate under the current CSI and fixed pilot signal.

As shown in Fig.~\ref{fig.Capacity-Net}, the input data of the Capacity-Net model consists of a set of transmitted pilot signals $\boldsymbol{Y}$ generated under fixed CSI with different phase shifts of the RIS, along with the specific RIS phase shift intended for estimation. The Capacity-Net model is designed to learn the intricate mapping between RIS phase shifts and achievable rates. Thus, the Capacity-Net model is trained to comprehend the relationship between the pilot signal under fixed CSI and the RIS precoder. 

The architecture of Capacity-Net employs a fully connected neural network, as depicted in Fig.~\ref{fig.Capacity-Net_neural}. Two input data of the Capacity-Net model are comprised of the pilot signal and RIS precoder. 
Subsequently, two input data are merged into a one-dimensional vector via the concatenate layer and fed into the Capacity-Net model for computing achievable rates.
The concatenate layer is used to concatenate multiple inputs together to provide a richer feature representation.
The estimated achievable rate is denoted as $\hat{R} \in \mathbb{R}$. The MSE is adopted as the loss function to measure the difference between the estimated achievable rate $\hat{R}$ and the anticipated achievable rate $R$. The proposed Capacity-Net model updates the weighting in the neural network until the convergence of the MSE loss function.

\subsection{Capacity-Net-Based Unsupervised Learning}

\begin{figure}[t]
\begin{center}
\includegraphics[width = 90mm]{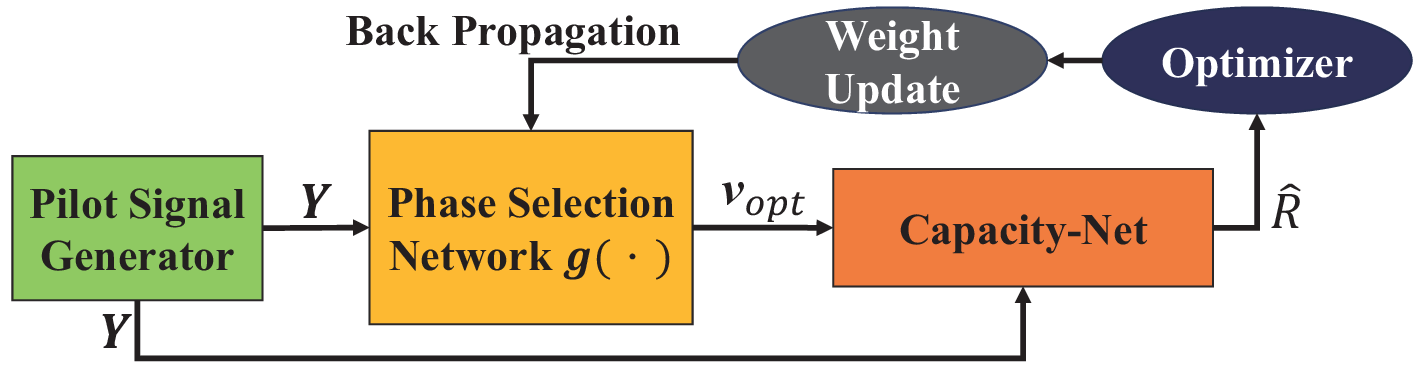}
\end{center}
\caption[c]{Capacity-Net-based unsupervised learning model training flow chart.}
\label{fig.Capacity-Net_Unsupervised}
\end{figure}

To eliminate the CSI requirement during the training phase, the Capacity-Net model is adopted instead of the computation of achievable rate in Fig.~\ref{fig.Unsupervised}. As shown in Fig.~\ref{fig.Capacity-Net_Unsupervised}, the proposed Capacity-Net-based unsupervised learning model does not need the CSI as the input data. Since the trained Capacity-Net model extracts the information of the channel from the pilot signal, it finds the RIS precoder which maximizes the achievable rate based on the pilot signal. Therefore, Capacity-Net-based unsupervised learning does not need to perform channel estimation to obtain perfect CSI.

\section{Simulation Results}
In this section, we demonstrate the resulting performance of the proposed method through computer simulations. 
We consider the MIMO system with $N_t=2$ transmit antennas and $N_r=6$ antenna assisted by a $4 \times 4$ RIS system.
It is assumed that there are $\Lambda=3$ NLOS paths for each link.
The Rician factor is set as $K=10$ and the AoAs and AoDs of all links are uniformly distributed over $[0, \pi]$. 
The path loss factors are $L_t=-72$ dB and $L_r=-66$ dB, and the noise variance is $-120$ dBm.

The hyperparameter settings of the fully connected neural networks of unsupervised learning neural network $g(\cdot)$ as shown in Tab.~\ref{tab.unsupervised_parameters}.
On the other hand, the hyperparameter settings of Capacity-Net are described as follows:
The fully connected architecture of Capacity-Net has an input layer with a dimension equal to the sum of the input and output size in $g(\cdot)$. The hidden layers consist of neural network nodes totaling 4096, with an output size of 1. The optimizer employed for the entire neural network is Adam, with a learning rate set to $10^{-5}$. The training set is $10^{6}$ and the testing set is $10\%$ of the training set.

\begin{table}[ht]
  \caption{Hyperparameter settings of unsupervised learning neural network $ g(\cdot) $.}
  \centering
  \begin{tabular}{|p{2.5cm}|c|}
    \hline
    \textbf{Parameter} & \textbf{Value} \\
    \hline
    Input layer& 384/768/1536/3072\\
    \hline
    Hidden layer& 4096\\\hline
 Output layer&16/36/64\\\hline
 Training set&$1 \times 10^6$\\\hline
 Testing set&$10^5$\\\hline
 Batch size&256\\\hline
 Learning rate&$3 \times 10^{-5}$\\\hline
 Optimizer&Adam\\\hline
  \end{tabular}
\label{tab.unsupervised_parameters}
\end{table}

We compare the proposed Capacity-Net model with three baselines: \textbf{(1) DSM~\cite{DSM}:} It uses the perfect CSI and iteratively performs the block coordinate descent to find the optimal phase shifts of RIS precoding until convergence. \textbf{(2) unsupervised learning:} In the training phase, it uses the pilot signals, phase shifts of RIS precoding, and perfect CSI to find the mapping that maximizes the achievable rate. In the testing phase, the pre-trained model does not need the perfect CSI. \textbf{(3) random selection:} The phase of RIS is randomly selected from the codebook.

\begin{figure}[t]
\begin{center}
\includegraphics[width = 90mm]{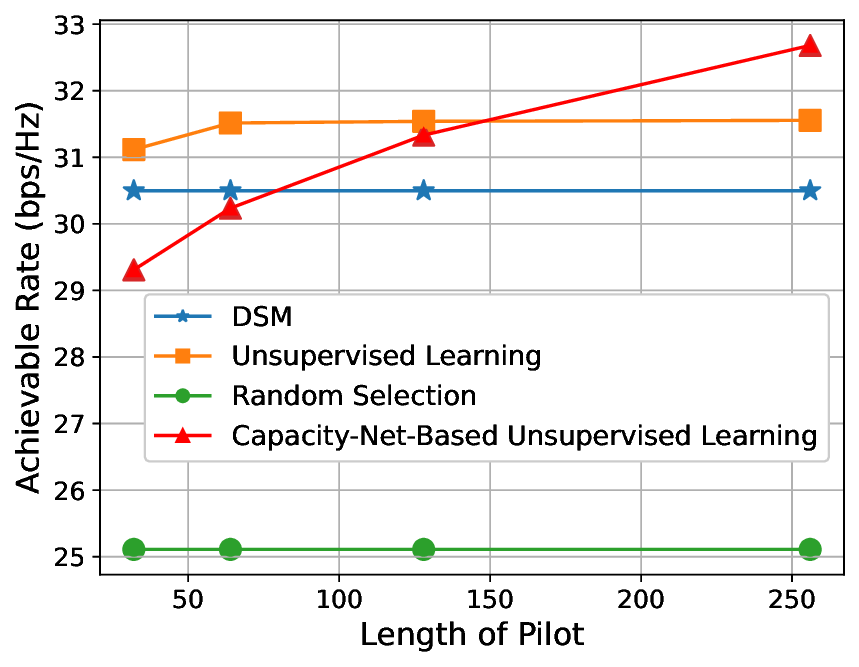}
\end{center}
\caption[c]{Achievable rates comparison for different lengths of pilot signal.}
\label{fig.rate_pilot}
\end{figure}

As shown in Fig.~\ref{fig.rate_pilot}, we compare the achievable rates of four methods in the different pilot signal lengths. The performances of DSM and random selection are not dependent on the length of the pilot signal. The unsupervised learning only slightly increases the achievable rate during the 40 to 60 number of pilot signals and no significant increase in the remaining pilot signal length range.
The proposed Capacity-Net model significantly increases the achievable rate with the length of the pilot signal.

\begin{figure}[t]
\begin{center}
\includegraphics[width = 90mm]{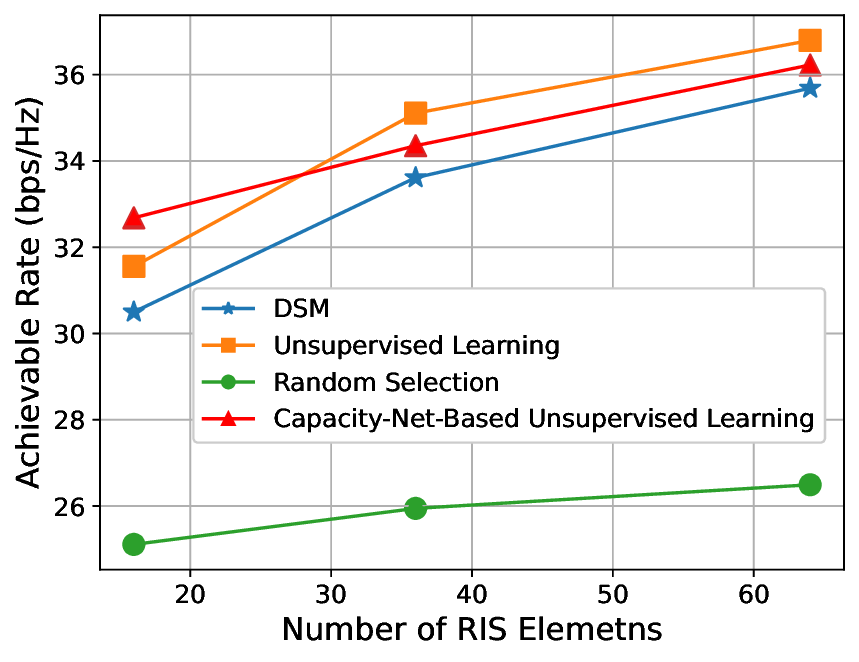}
\end{center}
\caption[c]{Achievable rates comparison for different numbers of RIS elements.}
\label{fig.rate_RIS}
\end{figure}

As shown in Fig.~\ref{fig.rate_RIS}, these four methods increase their achievable rates with the number of RIS elements. Since the number of RIS elements leads to the increased dimension of channel matrices, the unsupervised learning with perfect CSI outperforms the proposed Capacity-Net-based unsupervised learning without perfect CSI.
On the other hand, the unsupervised learning with/without the Capacity-Net model outperforms our previous work DSM.

\begin{figure}[t]
\begin{center}
\includegraphics[width = 90mm]{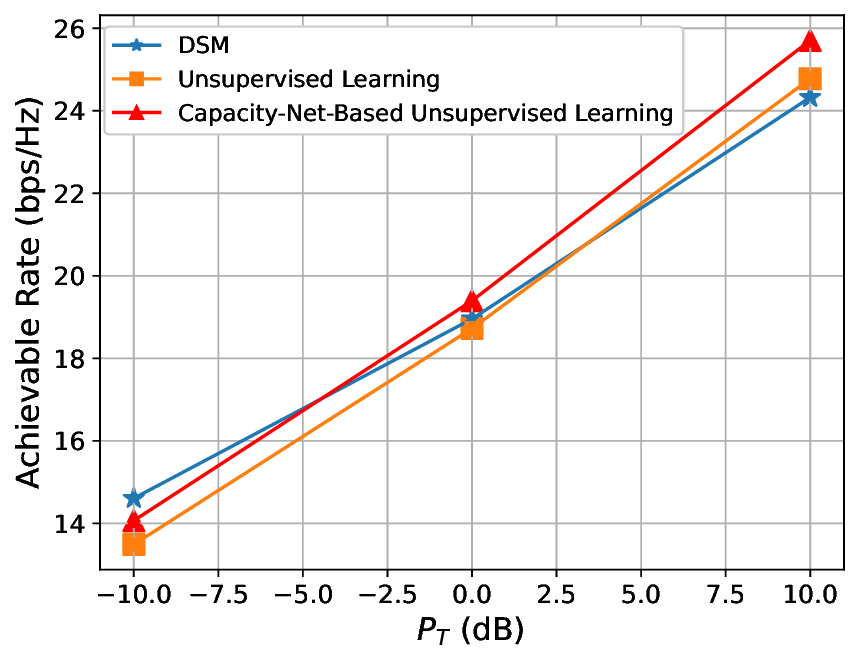}
\end{center}
\caption[c]{Generalizability for different transmitted power conditions.}
\label{fig.generalizability}
\end{figure}

As shown in Fig.~\ref{fig.generalizability}, we compare the generalizability of DSM, unsupervised learning, and Capacity-Net-based unsupervised learning.
In our simulation, the pilot signals are transmitted on the channel of SNR = 180 dB, the length of the pilot is set to $M = 16$, and the time slot is set to the same as the pilot length $L = 16$.
For different SNR conditions, Capacity-Net-based unsupervised learning fully outperforms unsupervised learning in terms of achievable rates by $4\%$.
When the SNR is less than 160 dB, unsupervised learning is gradually weaker than DSM in terms of achievable rate.
Therefore, this result demonstrates the remarkable generalization of Capacity-Net-based unsupervised learning.

\begin{figure}[t]
\begin{center}
\includegraphics[width = 90mm]{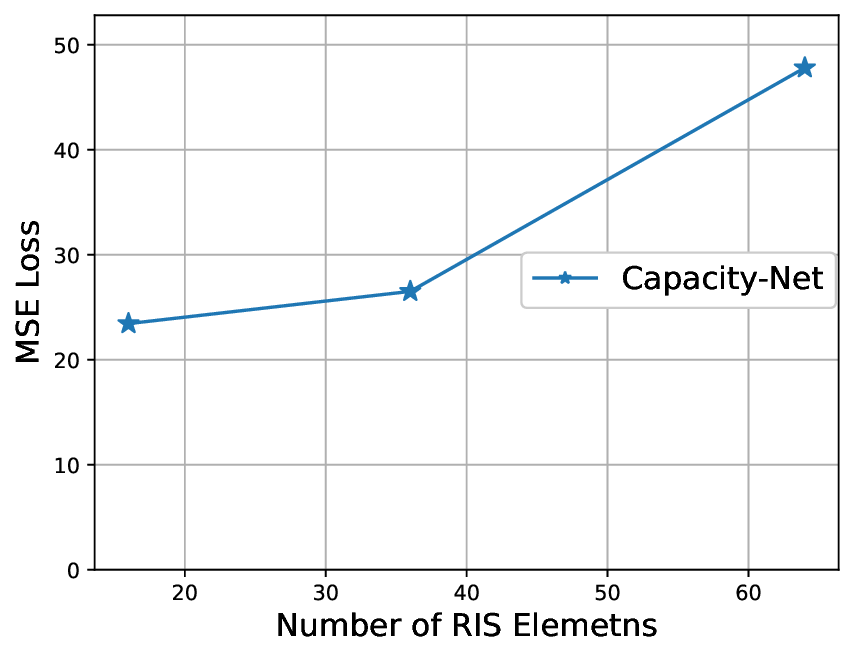}
\end{center}
\caption[c]{MSE loss for different numbers of RIS elements.}
\label{fig.MSE}
\end{figure}

As shown in Fig.~\ref{fig.MSE}, the number of RIS elements increased with the mean square error between the estimated achievable rate $\hat{R}$ and the anticipated achievable rate $R$.

\section{Conclusion}

This paper proposes Capacity-Net-based unsupervised learning to maximize the achievable rate for reflecting intelligent surface RIS-aided mmWave MIMO system.
First, we use the pilot signal as the input training data $\boldsymbol{Y}$ and adopt perfect CSI between transmitter-RIS and RIS-receiver to compute the achievable rate as the loss function to train unsupervised learning neural network  $g(\cdot)$.
Second, we design a Capacity-Net-based DL model to learn the mapping between RIS precoding vector $\boldsymbol{v}$ and the pilot signal $\boldsymbol{Y}$ from the knowledge of the achievable rate under the current perfect CSI.
Third, we adopt the trained Capacity-Net-based DL model to assist the unsupervised learning without perfect CSI for learning the mapping pilot signal and RIS precoding.
In the testing phase, our approach leveraged implicit channel estimation through received pilot signals to design the RIS precoding without prior channel estimation. 
Simulation results demonstrate the proposed Capacity-Net-based DL model outperforms unsupervised learning and the DSM algorithm. 
In addition, Capacity-Net-based training achieves high achievable rates with varying pilot lengths and RIS elements that show efficiency and robustness. 

For future research, we will jointly design the active precoder of the transmitter and the passive precoder of the RIS. Further, this work will be extended to the multi-RIS-aided multi-user MISO system and simultaneously transmitting and reflecting (STAR)-RIS-aided MIMO system.

\bibliographystyle{IEEEtran}

\end{document}